\documentclass[aip,jmp,preprint]{revtex4-1} 

\usepackage{amsmath}  
\usepackage{amsfonts} 
\usepackage{graphicx} 
\usepackage{url}

\usepackage[bottom]{footmisc}
\usepackage{amssymb}
\usepackage{booktabs}
\usepackage[format=plain,labelsep=period]{caption}
\usepackage[labelformat=simple]{subcaption}
\usepackage{caption}
\usepackage{textcomp}
\usepackage{mathrsfs}
\usepackage[shortlabels]{enumitem}
\usepackage{upgreek}
\usepackage{fullpage}
\usepackage[title]{appendix}
\usepackage[pagestyles]{titlesec}

\newcommand {\dd}[3][]{\frac{d^{#1}#2}{d#3^{#1}}}

\begin{document}

\title{Fourth-order dynamics of the damped harmonic oscillator}


\author{John W. Sanders\\
           Department of Mechanical Engineering \\
	California State University, Fullerton \\
	Fullerton, CA 92831, United States \\
	\email{jwsanders@fullerton.edu} \\
	ORCiD: 0000-0003-3059-3815}



\date{Received: date / Accepted: date}

\maketitle


\vspace{12pt}

\section*{Abstract}

It is shown that the classical damped harmonic oscillator belongs to the family of fourth-order Pais-Uhlenbeck oscillators. It follows that the solutions to the damped harmonic oscillator equation make the Pais-Uhlenbeck action stationary. Two systematic approaches are given for deriving the Pais-Uhlenbeck action from the damped harmonic oscillator equation, and it may be possible to use these methods to identify stationary action principles for other dissipative systems which do not conform to Hamilton's principle. It is also shown that for every damped harmonic oscillator $x$, there exists a two-parameter family of dual oscillators $y$ satisfying the Pais-Uhlenbeck equation. The damped harmonic oscillator and any of its duals can be interpreted as a system of two coupled oscillators with atypical spring stiffnesses (not necessarily positive and real-valued). For overdamped systems, the resulting coupled oscillators should be physically achievable and may have engineering applications. Finally, a new physical interpretation is given for the optimal damping ratio $\zeta=1/\sqrt{2}$ in control theory.

\section{Introduction and literature review}

Ever since Newton~\cite{Newton1687} first laid the foundation for mechanics as a formal system, the development of analytical mechanics has progressed via the principle of virtual work and its generalization, d'Alembert's principle,~\cite{dAlembert1743} reaching its pinnacle in the principle of stationary action.~\cite{Lagrange1811,Hamilton1834,Hamilton1835} During any time interval $t\in(t_{1},t_{2})$, the actual motion of a conservative mechanical system is given by a critical point of the action functional
\begin{equation}
\mathcal{S}=\int_{t_{1}}^{t_{2}}Ldt,
\end{equation}
where the Lagrangian $L=T-V$ is the difference between the system's total kinetic energy $T$ and total potential energy $V$. Put simply, the action is stationary for the actual motion of a conservative system:
\begin{equation}\label{eq:Hamilton}
\delta\mathcal{S}=0.
\end{equation}
While sometimes attributed to Hamilton,~\cite{Hamilton1834,Hamilton1835} the principle of stationary action was already well known by Lagrange~\cite{Lagrange1811} and contemporaries.~\cite{Lanczos1949,Gray2007} It has been demonstrated that the action always achieves either a local minimum or a saddle point, but never a local maximum.~\cite{Jacobi1837,Whittaker,Gray2007} 

When dissipative forces are present, Hamilton's principle \eqref{eq:Hamilton} may be exchanged for the d'Alembert-Lagrange principle~\cite{Lagrange1811,Kane2000,Baddour2007,Flannery2011}
\begin{equation}\label{eq:dAlembertLagrange}
\delta \mathcal{S}+\int_{t_{1}}^{t_{2}}Q_{i}\delta{q}_{i}dt=0,
\end{equation}
where the $q_{i}$ are the generalized coordinates and the $Q_{i}$ are the generalized dissipative forces. Like d'Alembert's principle,~\cite{dAlembert1743} Hamilton's principle \eqref{eq:Hamilton} and the d'Alembert-Lagrange principle \eqref{eq:dAlembertLagrange} can be used as the basis for numerical solution methods.~\cite{Gray1996,Kim2013} Indeed, much modern work has been devoted to the development of accurate and efficient variational time integrators based on these principles.~\cite{Kane2000,Marsden2001,Lew2003,Lew2004,Lew2004a,Kale2007,OberBloebaum2011,Kraus2015,Hall2015,Lew2016,Schmitt2018,Man2020,Limebeer2020,Kraus2021} 
 Nevertheless, the fact remains that the d'Alembert-Lagrange principle is not a true variational principle, because in general the virtual work terms $Q_{i}\delta{q}_{i}$ are not the exact variations of a work function.~\cite{Lanczos1949} As a result, those terms must be inserted into the equation manually \emph{after} the variation operator $\delta$ has been applied to the action. The \emph{ad hoc} insertion of the dissipative terms into an otherwise variational principle is extremely dissatisfying on physical, mathematical, and aesthetic grounds, and over the last two centuries many attempts have been made to extend the Lagrangian framework and the stationary action principle to dissipative systems. 
 
 \begin{figure}[b]
 \centering
 \includegraphics[width=0.4\textwidth]{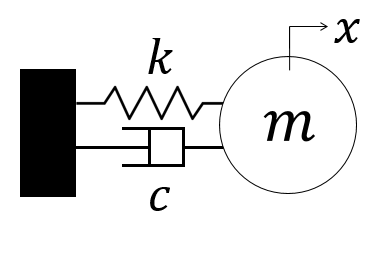}
 \caption{The damped harmonic oscillator.}
 \label{fig:DHO}
 \end{figure}

One of the most ubiquitous models of a dissipative system is given by the damped harmonic oscillator, as shown in Figure~\ref{fig:DHO}. It consists of a mass $m$ attached to a rigid support by a linear spring of stiffness $k$ and a linear damper of damping coefficient $c$, with $x(t)$ the generalized coordinate along the single degree of freedom. The governing equation of motion is given by
\begin{equation}\label{eq:DHOequation0}
m\ddot{x}+c\dot{x}+kx=0,
\end{equation}
or equivalently,
\begin{equation}\label{eq:DHOequation}
\ddot{x}+2\omega\zeta\dot{x}+\omega^{2}x=0,
\end{equation}
where we have introduced the natural frequency $\omega=\sqrt{k/m}$ and damping ratio $\zeta=c/2m\omega$. Henceforth, we shall regard $\omega>0$ and $\zeta\geq0$ as real-valued model parameters. Depending on the damping ratio, the damped harmonic oscillator exhibits different behaviors, including the absence of damping ($\zeta=0$), underdamping ($0<\zeta<1$), critical damping ($\zeta=1$), and overdamping ($\zeta>1$). While the damped harmonic oscillator does not conform to Hamilton's principle \eqref{eq:Hamilton}, there are known variational principles that reproduce the governing equation \eqref{eq:DHOequation}. As discussed in detail by Limebeer \emph{et al.},~\cite{Limebeer2020} one class of stationary action principles is based on non-standard Lagrangians. For example, the dual system method of Bateman, Morse, Feshbach, and Tikochinsky~\cite{Bateman1931,Morse1953,Feshbach1977} employs a Lagrangian of the form
\begin{equation}
L=\dot{x}\dot{y}+\omega\zeta(x\dot{y}-y\dot{x})-\omega^{2}xy,
\end{equation}
where $x$ and $y$ are both to be varied. The Euler-Lagrange equations yield two oscillators: the damped harmonic oscillator \eqref{eq:DHOequation} and a dual anti-damped oscillator
\begin{equation}
\ddot{y}-2\omega\zeta\dot{y}+\omega^{2}y=0.
\end{equation}
Another example is given by the Kanai-Caldirola oscillator,~\cite{Caldirola1941,Kanai1948} which uses a Lagrangian of the form
\begin{equation}
L=\frac{1}{2}e^{2\omega\zeta t}(\dot{x}^{2}-\omega^{2}x^{2}),
\end{equation}
with Euler-Lagrange equation
\begin{equation}
e^{2\omega\zeta t}(\ddot{x}+2\omega\zeta\dot{x}+\omega^{2}x)=0.
\end{equation}
A final approach that will be mentioned here is the Caldeira-Leggett model,~\cite{Caldeira1983,Zwanzig1973,Ford1987} which treats a system as embedded within a heat bath, so that the combination of the system and its surroundings is conservative.~\cite{Limebeer2020} An example is provided by Fukagawa and Fujitani,~\cite{Fukagawa2012} who used Hamilton's principle coupled with a nonholonomic constraint on the entropy to incorporate the damping force into the damped harmonic oscillator equation as well as the viscous terms into the momentum balance equation for a viscous fluid, one special case being the Navier-Stokes equations.~\cite{Stokes1845}

Quite independent of the damped harmonic oscillator is the Pais-Uhlenbeck oscillator,~\cite{Pais1950} which is used as a representative ``toy model''  to study higher-derivative theories for quantum gravity.~\cite{Mostafazadeh2010,Baleanu2012} In general, the Pais-Uhlenbeck oscillator of order $2n$ is given by
\begin{equation}
\prod_{j=1}^{n}\left(\dd[2]{}{t}+\omega_{j}^{2}\right)x=0,
\end{equation}
where the $\omega_{j}$ are real-valued parameters. For $n=2$, this yields the fourth-order equation
\begin{equation}
\ddddot{x}+(\omega_{1}^{2}+\omega_{2}^{2})\ddot{x}+\omega_{1}^{2}\omega_{2}^{2}x=0.
\end{equation}
In the present work, we will establish that the damped harmonic oscillator belongs to the family of fourth-order Pais-Uhlenbeck oscillators. To date, it seems that the connection between the damped harmonic oscillator and the Pais-Uhlenbeck oscillator has gone unnoticed. This connection leads to two interesting conclusions: (i) that the damped harmonic oscillator is a critical point of the Pais-Uhlenbeck action, and (ii) that for every damped harmonic oscillator $x$, there exists a two-parameter family of coupled dual oscillators $y$.

The remainder of this paper is organized as follows. In Section~\ref{sec:DHOisPU}, we show that every solution to the damped harmonic oscillator equation \eqref{eq:DHOequation} is also a solution to a fourth-order equation of the Pais-Uhlenbeck type. In Section~\ref{sec:coupledsystem}, we investigate the two-parameter family of coupled-oscillator systems, leading to new interpretations for the various damping regimes. In Section~\ref{sec:optimalcontrol}, we examine the special case of $\zeta=1/\sqrt{2}$, which leads to new insights into this ``optimal control value.'' In Section~\ref{sec:variationalformulation}, we discuss the corresponding variational formulation based on the Pais-Uhlenbeck action. We also present two systematic methods for arriving at the Pais-Uhlenbeck action from the original damped harmonic oscillator equation, which may help identify variational principles for other dissipative systems. Finally, in Section~\ref{sec:conclusion}, we conclude with a brief summary of the present results.

\section{The damped harmonic oscillator as a Pais-Uhlenbeck oscillator}\label{sec:DHOisPU}

We begin by observing that every solution to \eqref{eq:DHOequation} is also a solution to the following fourth-order equation:
\begin{equation}\label{eq:PUequation}
\ddddot{x}+4\omega^{2}\left(\frac{1}{2}-\zeta^{2}\right)\ddot{x}+\omega^{4}x=0,
\end{equation}
which is a kind of Pais-Uhlenbeck equation in which the coefficient of $\ddot{x}$ may be positive, negative, or zero, depending on the damping ratio. This can be demonstrated as follows. Differentiating \eqref{eq:DHOequation} twice, we obtain
\begin{equation}\label{eq:secondderivative}
\ddddot{x}+2\omega\zeta\dddot{x}+\omega^{2}\ddot{x}=0.
\end{equation}
Substituting
\begin{equation}
\dddot{x}=-2\omega\zeta\ddot{x}-\omega^{2}\dot{x}
\end{equation}
and
\begin{equation}
\dot{x}=-\frac{1}{2\omega\zeta}(\ddot{x}+\omega^{2}x)
\end{equation}
into \eqref{eq:secondderivative} yields \eqref{eq:PUequation}. We conclude that the damped harmonic oscillator belongs to the family of Pais-Uhlenbeck oscillators. Indeed, with the same initial conditions
\begin{align}
x(0) &= x_{0}, \\
\dot{x}(0) &= v_{0}, \\
\ddot{x}(0) &= a_{0} \equiv -2\omega\zeta v_{0}-\omega^{2}x_{0}, \\
\dddot{x}(0) &= j_{0} \equiv -2\omega\zeta a_{0}-\omega^{2}v_{0},
\end{align}
the uniqueness theorem guarantees that the Pais-Uhlenbeck oscillator \eqref{eq:PUequation} will yield the same solution as the damped harmonic oscillator \eqref{eq:DHOequation}.

Using a trial solution of the form
\begin{equation}
x(t)=Ae^{\lambda t},
\end{equation}
we solve the characteristic polynomial
\begin{equation}
\lambda^{4}+4\omega^{2}\left(\frac{1}{2}-\zeta^{2}\right)\lambda^{2}+\omega^{4}=0,
\end{equation}
yielding four eigenvalues:
\begin{align}
\lambda_{1,2}=&\pm\omega\sqrt{-2\left(\frac{1}{2}-\zeta^{2}\right)+\sqrt{4\left(\frac{1}{2}-\zeta^{2}\right)^{2}-1}}=\pm\omega(\zeta+\sqrt{\zeta^{2}-1}), \\
\lambda_{3,4}=&\pm\omega\sqrt{-2\left(\frac{1}{2}-\zeta^{2}\right)-\sqrt{4\left(\frac{1}{2}-\zeta^{2}\right)^{2}-1}}=\pm\omega(\zeta-\sqrt{\zeta^{2}-1}).
\end{align}
The behavior of the solution is determined by the damping coefficient. Just as in the classical theory of the damped harmonic oscillator, there are four cases:
\begin{enumerate}[1.] 
  \item \textbf{No damping, $\zeta=0$:} For undamped oscillators, $\lambda_{1}=\lambda_{4}=+i\omega$ and $\lambda_{2}=\lambda_{3}=-i\omega$. In this case, we have
  \begin{equation}
  x(t)=Ae^{+i\omega t}+Bte^{+i\omega t}+Ce^{-i\omega t}+Dte^{-i\omega t}.
  \end{equation}
  This reduces to the classical simple harmonic oscillator solution
  \begin{equation}
  x(t)=E\cos{(\omega t)}+F\sin{(\omega t)}
  \end{equation}
  upon setting $B=D=0$, $A+C=E$, and $i(A-C)=F$.
  \item \textbf{Underdamping, $0<\zeta<1$:} In this case, the eigenvalues are complex-valued and distinct. Writing
  \begin{align}
  \lambda_{1,2}=&\pm\omega(\zeta+i\sqrt{1-\zeta^{2}}), \\
  \lambda_{3,4}=&\pm\omega(\zeta-i\sqrt{1-\zeta^{2}}),
  \end{align}
  where $i$ is the imaginary unit, we have
  \begin{equation}
  x(t)=Ae^{\omega\zeta t}e^{i\omega\sqrt{1-\zeta^{2}}t}+Be^{-\omega\zeta t}e^{-i\omega\sqrt{1-\zeta^{2}}t}+Ce^{\omega\zeta t}e^{-i\omega\sqrt{1-\zeta^{2}}t}+De^{-\omega\zeta t}e^{i\omega\sqrt{1-\zeta^{2}}t}.
  \end{equation}
  This reduces to the classical damped solution
  \begin{equation}
  x(t)=e^{-\omega\zeta t}\left[E\cos{\left(\omega\sqrt{1-\zeta^{2}}t\right)}+F\sin{\left(\omega\sqrt{1-\zeta^{2}}t\right)}\right].
  \end{equation}
  upon setting $A=C=0$, $B+D=E$, and $-i(B-D)=F$. 
  \item \textbf{Critical damping, $\zeta=1$:} For critically damped systems, $\lambda_{1}=\lambda_{3}=+\omega$ and $\lambda_{2}=\lambda_{4}=-\omega$. In this case, we have
  \begin{equation}
  x(t)=Ae^{+\omega t}+Bte^{+\omega t}+Ce^{-\omega t}+Dte^{-\omega t}.
  \end{equation}
  This reduces to the classical damped harmonic solution
  \begin{equation}
  x(t)=(C+Dt)e^{-\omega t}
  \end{equation}
  upon setting $A=B=0$.
  \item \textbf{Overdamping, $\zeta>1$:} For overdamped systems, all eigenvalues are real-valued and distinct, leading to a linear combination of exponential growth and decay:
  \begin{equation}
  x(t)=Ae^{\lambda_{1}t}+Be^{\lambda_{2}t}+Ce^{\lambda_{3}t}+De^{\lambda_{4}t}.
  \end{equation}
  This reduces to the classical damped harmonic solution
  \begin{equation}
  x(t)=Be^{-\omega(\zeta+\sqrt{\zeta^{2}-1})t}+De^{-\omega(\zeta-\sqrt{\zeta^{2}-1})t}
  \end{equation}
  upon setting $A=C=0$.
\end{enumerate}

\section{Equivalent coupled-oscillator system}\label{sec:coupledsystem}

It is well-known that the fourth-order Pais-Uhlenbeck equation \eqref{eq:PUequation} is mathematically equivalent to the following system of two second-order equations:
\begin{align}
\ddot{x}+\mu_{1}x-\rho_{1}y=&0, \\
\ddot{y}+\mu_{2}y-\rho_{2}x=&0,
\end{align}
 \begin{figure}[t]
 \centering
 \includegraphics[width=0.9\textwidth]{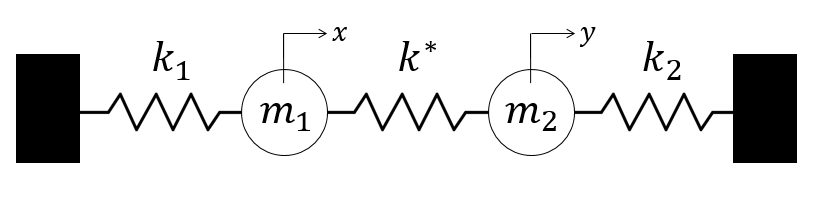}
 \caption{Coupled oscillator system equivalent to the fourth-order Pais-Uhlenbeck oscillator.}
 \label{fig:coupled}
 \end{figure}
where
\begin{equation}\label{eq:aux1}
\mu_{1}+\mu_{2}=4\omega^{2}\left(\frac{1}{2}-\zeta^{2}\right),
\end{equation}
\begin{equation}\label{eq:aux2}
\mu_{1}\mu_{2}-\rho_{1}\rho_{2}=\omega^{4}.
\end{equation}
and both $x(t)$ and $y(t)$ satisfy the Pais-Uhlenbeck equation \eqref{eq:PUequation} individually. This can be thought of as a system of two coupled oscillators, as shown schematically in Figure~\ref{fig:coupled}. The springs in Figure~\ref{fig:coupled} are not meant to suggest that a damped oscillator is equivalent to two \emph{conservative} oscillators: as we will see, the stiffnesses of the supporting springs ($k_{1}$, $k_{2}$) and coupling spring ($k^{*}$) are generally not real-valued and positive, and so the dynamics are not conservative. In general, the masses and stiffnesses of the coupled system are related to the model parameters as follows:
\begin{equation}
\mu_{1}=\frac{k_{1}+k^{*}}{m_{1}}, \quad \rho_{1}=\frac{k^{*}}{m_{1}}, \quad \mu_{2}=\frac{k^{*}+k_{2}}{m_{2}}, \quad \rho_{2}=\frac{k^{*}}{m_{2}}.
\end{equation}
whence we obtain
\begin{equation}
\frac{m_{2}}{m_{1}}=\frac{\rho_{1}}{\rho_{2}}, \quad \frac{k_{1}}{k^{*}}=\frac{\mu_{1}}{\rho_{1}}-1, \quad \frac{k_{2}}{k^{*}}=\frac{\mu_{2}}{\rho_{2}}-1.
\end{equation}
With initial conditions
\begin{align}
x(0) &= x_{0}, &y(0) = (a_{0}+\mu_{1}x_{0})/\rho_{1}, \\
\dot{x}(0) &= v_{0}, &\dot{y}(0) = (j_{0}+\mu_{1}v_{0})/\rho_{1},
\end{align}
$x$ will coincide with the damped harmonic oscillator. Now we observe that there are only two damped harmonic oscillator parameters ($\omega$ and $\zeta$), but there are four coupled oscillator parameters ($\mu_{1}$, $\mu_{2}$, $\rho_{1}$, and $\rho_{2}$). It follows that for every damped harmonic oscillator $x$ there exists a \emph{two-parameter family of dual oscillators} $y$ satisfying the Pais-Uhlenbeck equation. For example, provided $\zeta\neq1/\sqrt{2}$ (this is a unique case that will be treated in Section~\ref{sec:optimalcontrol}), we may set $\mu_{2}=(1/\alpha-1)\mu_{1}$ ($\alpha\neq0$, $\alpha\neq1$) and $\rho_{2}=\beta^{-2}\rho_{1}$ ($\beta\neq0$), leading to
\begin{align}
\mu_{1}&=4\alpha\omega^{2}\left(\frac{1}{2}-\zeta^{2}\right), & \mu_{2}&=4(1-\alpha)\omega^{2}\left(\frac{1}{2}-\zeta^{2}\right),\\
\rho_{1}&=\beta\omega^{2}\sqrt{16\alpha(1-\alpha)\left(\frac{1}{2}-\zeta^{2}\right)^{2}-1}, & \rho_{2}&=\frac{1}{\beta}\omega^{2}\sqrt{16\alpha(1-\alpha)\left(\frac{1}{2}-\zeta^{2}\right)^{2}-1},
\end{align}
where we have chosen the positive radical for $\rho_{1}$ in order to make the coupling stiffness $k^{*}$ positive when $k^{*}$ is real-valued and $\beta$ is positive.

\subsection{Symmetric coupled oscillator}\label{sec:symmetric}

\begin{table}[b]
\caption{Summary of all possible scenarios for a symmetric coupled oscillator ($m_{1}=m_{2}=m\in\mathbb{R}$, $k_{1}=k_{2}=k$). The sign of $\rho$ has been chosen so as to make the coupling stiffness $k^{*}$ positive when it is real-valued.}
\label{tab:cases}
\centering
\begin{tabular}{lccccc}
\hline
Case & $\mu$ & $\rho$ & $k^{*}=m\rho$ & $k=m\mu-k^{*}$ & $r=k^{*}/k$ \\
\hline
$\zeta=0$ & $+\omega^{2}$ & 0 & 0 & $+m\omega^{2}$ & 0 \\
$\zeta\in(0,1/\sqrt{2})$ & real, positive & imaginary & imaginary & complex & complex \\
$\zeta=1/\sqrt{2}$  & 0 & $+i\omega^{2}$ & $+im\omega^{2}$ & $-im\omega^{2}$ & -1 \\
$\zeta\in(1/\sqrt{2},1)$ & real, negative & imaginary & imaginary  & complex & complex \\
$\zeta=1$ & $-\omega^{2}$ & 0 & 0 & $-m\omega^{2}$ & 0  \\
$\zeta>1$  & real, negative & real, positive & real, positive & real, negative & real, negative \\
$\zeta\rightarrow+\infty$  & $-\infty$ & $+\infty$ & $+\infty$ & $-\infty$ & $-1/2$ \\
\hline
\end{tabular}
\end{table}

\begin{figure}[t]
 \centering
 \includegraphics[width=0.7\textwidth]{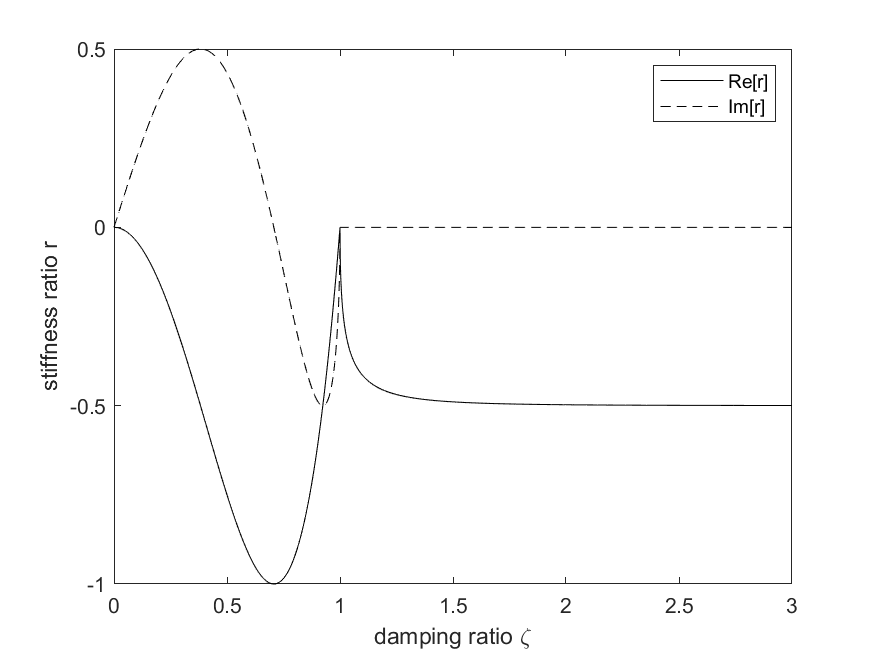}
 \caption{Real and imaginary parts of the stiffness ratio $r$ versus damping ratio $\zeta$.}
 \label{fig:stiffnessratio}
 \end{figure}

Henceforth, we will restrict attention to the symmetric coupled oscillator, in which the two masses are identical and real-valued ($m_{1}=m_{2}=m\in\mathbb{R}$) and the two supporting springs have identical stiffnesses ($k_{1}=k_{2}=k$, not necessarily real or positive), giving $\rho_{1}=\rho_{2}=\rho$ and $\mu_{1}=\mu_{2}=\mu$. This corresponds to the particular dual oscillator with $\alpha=1/2$ and $\beta=1$, giving
\begin{equation}
\mu=2\omega^{2}\left(\frac{1}{2}-\zeta^{2}\right),
\end{equation}
\begin{equation}
\rho=\omega^{2}\sqrt{4\left(\frac{1}{2}-\zeta^{2}\right)^{2}-1}=2\omega^{2}\zeta\sqrt{\zeta^{2}-1}.
\end{equation}
The coupling stiffness is given by
\begin{equation}
k^{*}=m\rho=2m\omega^{2}\zeta\sqrt{\zeta^{2}-1},
\end{equation}
and the supporting stiffnesses are given by
\begin{equation}
k=m\mu-k^{*}=m(\mu-\rho)=2m\omega^{2}\left(\frac{1}{2}-\zeta^{2}-\zeta\sqrt{\zeta^{2}-1}\right).
\end{equation}
We define the stiffness ratio as
\begin{equation}\label{eq:stiffnessratio}
r=\frac{k^{*}}{k}=\left(\frac{\mu}{\rho}-1\right)^{-1}=-\left[\frac{\zeta^{2}-\frac{1}{2}}{\zeta\sqrt{\zeta^{2}-1}}+1\right]^{-1}.
\end{equation}
The real and imaginary parts of the stiffness ratio $r$ are plotted versus $\zeta$ in Figure~\ref{fig:stiffnessratio}. Below we consider the various damping cases in turn, which are summarized in Table~\ref{tab:cases}.
\begin{enumerate}[1.]
  \item \textbf{No damping, $\zeta=0$:} In the absence of damping, $\mu=+\omega^{2}$, $\rho=0$, $k^{*}=0$, $k=+m\omega^{2}$, and $r=0$. This corresponds to a case of two uncoupled simple harmonic oscillators with positive stiffnesses and identical natural frequencies $\omega$.
  \item \textbf{Underdamping, $0<\zeta<1$:} For underdamping, the stiffnesses $k$ and $k^{*}$ are non-real. To be more precise, we must make a distinction between damping ratios less than, equal to, and greater than $\zeta=1/\sqrt{2}$:
  \begin{enumerate}[(a)]
    \item \textit{$0<\zeta<1/\sqrt{2}$}: We have that $\mu$ is real and positive, $\rho$ is imaginary, $k^{*}$ is imaginary, $k$ is complex, and the stiffness ratio $r$ is complex.
    \item \textit{$\zeta=1/\sqrt{2}$}: We have that $\mu=0$, $\rho=+i\omega^{2}$, $k^{*}=+im\omega^{2}$, $k=-im\omega^{2}$, and $r=-1$. Again, this case will be discussed in more detail in Section~\ref{sec:optimalcontrol}.
    \item \textit{$1/\sqrt{2}<\zeta<1$}: We have that $\mu$ is real and negative, $\rho$ is imaginary, $k^{*}$ is imaginary, $k$ is complex, and the stiffness ratio $r$ is complex.
  \end{enumerate}
  Figure~\ref{fig:xandyunderdamped} shows representative coupled oscillator dynamics for an underdamped case with $\zeta=0.2$, $\omega=1$, $x_{0}=1$, and $v_{0}=0$. In this case, the damped harmonic oscillator $x(t)$ is real-valued, while the dual oscillator $y(t)$ is imaginary.
 \begin{figure}[t]
 \centering
 \includegraphics[width=0.7\textwidth]{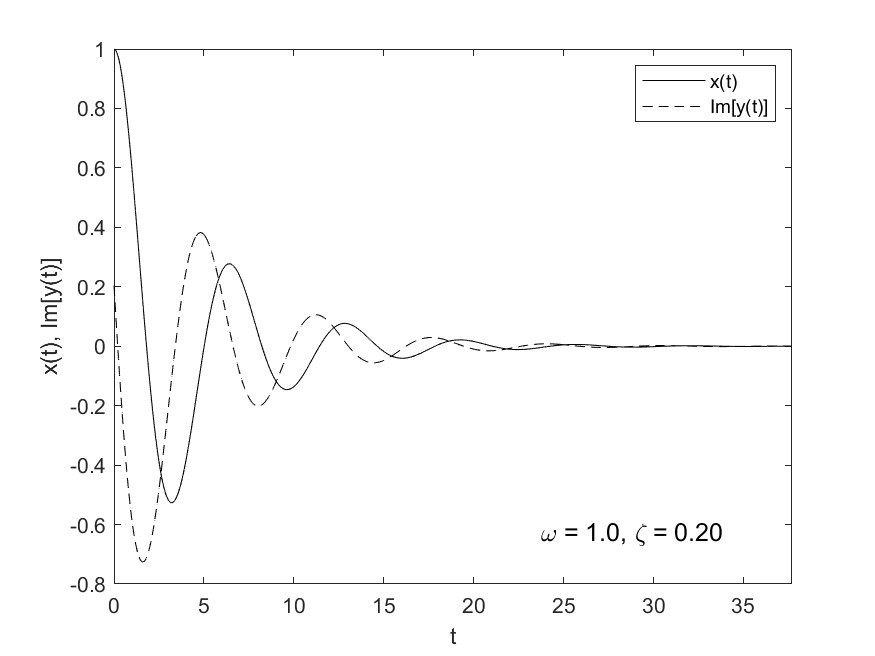}
 \caption{Representative coupled oscillator dynamics for an underdamped system with $\zeta=0.2$, $\omega=1$, $x_{0}=1$, and $v_{0}=0$. The damped harmonic oscillator $x(t)$ is real-valued, while the dual oscillator $y(t)$ is imaginary.}
 \label{fig:xandyunderdamped}
 \end{figure} 
  \item \textbf{Critical damping, $\zeta=1$:} For critical damping, $\mu=-\omega^{2}$, $\rho=0$, $k^{*}=0$, $k=-m\omega^{2}$, and $r=0$. As in the absence of damping, this corresponds to another case in which the coupling spring stiffness is zero. The difference is that here the supporting spring stiffnesses are \emph{negative}.
  \item \textbf{Overdamping, $\zeta>1$:} All quantities are real-valued. We have that $\mu$ is negative, $\rho$ is positive, $k^{*}$ is positive, $k$ is negative, and $r$ is negative. In the limit as $\zeta\rightarrow+\infty$, $k^{*}\rightarrow+\infty$ and $k\rightarrow-\infty$, and $r\rightarrow-1/2$. This leads to an interesting interpretation of overdamping. As the system becomes more and more overdamped, the coupling spring becomes ever more positively stiff, and the supporting springs become ever more negatively stiff, with their ratio approaching $-1/2$ asymptotically. Figure~\ref{fig:xandyoverdamped} shows representative coupled oscillator dynamics for an overdamped case with $\zeta=1.5$, $\omega=1$, $x_{0}=1$, and $v_{0}=0$. In this case, the damped harmonic oscillator $x(t)$ and the dual oscillator $y(t)$ are both real-valued. 
 \begin{figure}[b]
 \centering
 \includegraphics[width=0.7\textwidth]{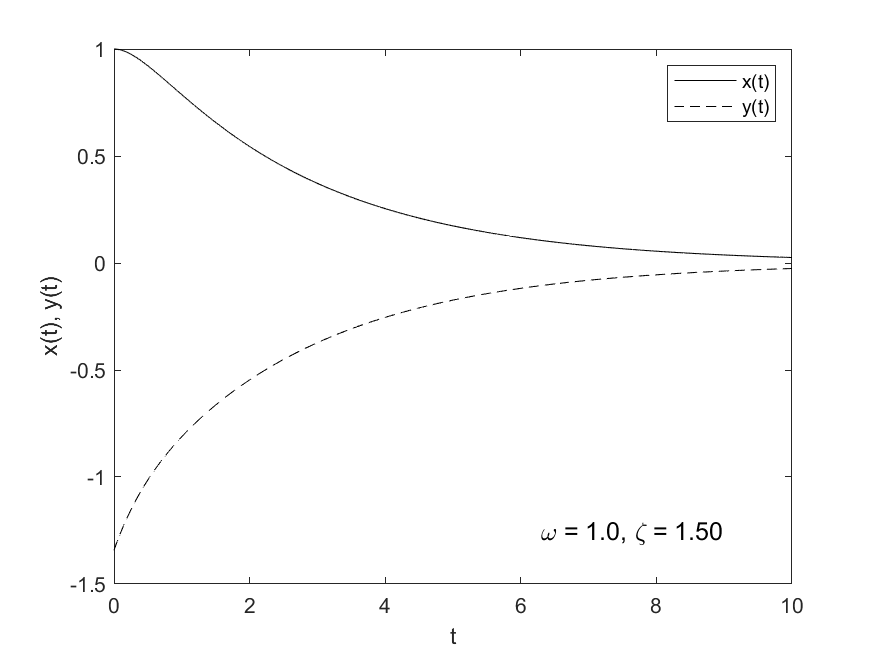}
 \caption{Representative coupled oscillator dynamics for an overdamped system with $\zeta=1.5$, $\omega=1$, $x_{0}=1$, and $v_{0}=0$. The damped harmonic oscillator $x(t)$ and the dual oscillator $y(t)$ are both real-valued.}
 \label{fig:xandyoverdamped}
 \end{figure}
\end{enumerate}
It should be noted that spring-like elements with effectively negative stiffnesses are physically achievable.~\cite{Lee2007} Thus, it should be possible to test the results presented here for overdamped oscillators experimentally. Coupled oscillators mimicking the overdamped harmonic oscillator may even find applications to vibration suppression systems.
 

\section{Optimal control: $\zeta=1/\sqrt{2}$}\label{sec:optimalcontrol}
 
 \begin{figure}[b]
 \centering
 \includegraphics[width=0.7\textwidth]{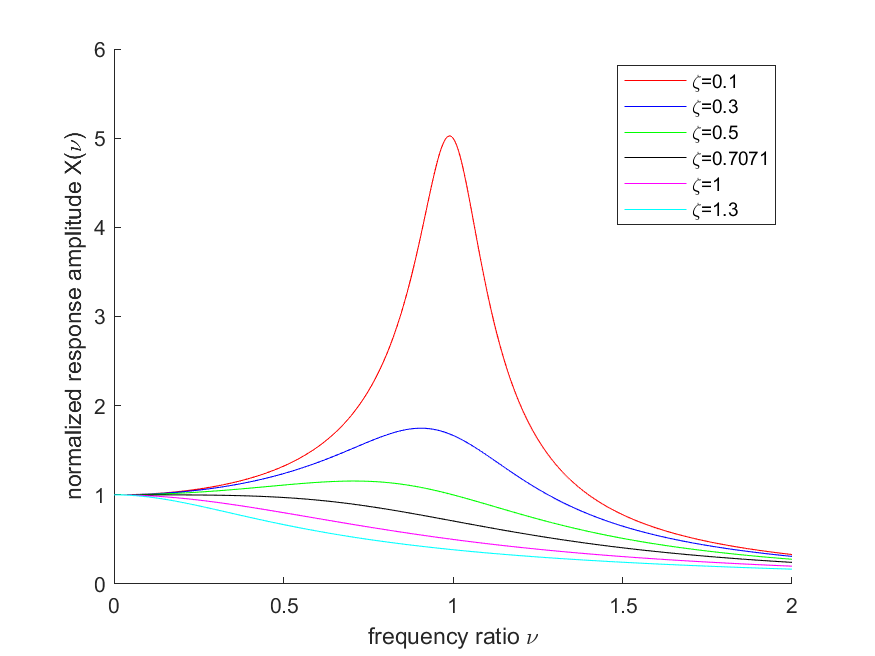}
 \caption{Plot of normalized response amplitude $X(\nu)$ versus frequency ratio $\nu$ for various damping ratios $\zeta$. The local maximum does not exist for $\zeta>1/\sqrt{2}\approx0.7071$.}
 \label{fig:bode}
 \end{figure}

Consider a forced damped harmonic oscillator of the form
\begin{equation}
\ddot{x}+2\omega\zeta\dot{x}+\omega^{2}x=f_{0}\sin{(\Upomega t)},
\end{equation}
where $f_{0}$ is the forcing magnitude and $\Upomega$ is the forcing frequency. At steady state, the solution has normalized magnitude
\begin{equation}
X(\nu)=\frac{1}{\sqrt{(1-\nu^{2})^{2}+(2\zeta \nu)^{2}}},
\end{equation}
where $\nu=\Upomega/\omega$ is the frequency ratio. A plot of $X(\nu)$ versus $\nu$ for various values of $\zeta$ is shown in Figure~\ref{fig:bode}. It is straightforward to show that $X(\nu)$ has a local maximum at $\nu=\sqrt{1-2\zeta^{2}}$, provided $\zeta<1/\sqrt{2}$. For $\zeta>1/\sqrt{2}$, there is no longer a local maximum. The critical value $\zeta=1/\sqrt{2}$ at which the local maximum disappears happens to be the ``optimal control value'' in terms of the tradeoff between rise time and percent overshoot in response to a step input.~\cite{Jacobs1965} 

Looking at the second-order damped harmonic oscillator equation \eqref{eq:DHOequation}, $\zeta=1/\sqrt{2}$ does not appear to have a special role; it is simply one case of underdamping. However, it is \emph{immediately} clear from the Pais-Uhlenbeck equation \eqref{eq:PUequation} that $\zeta=1/\sqrt{2}$ is significant. The inertial term,
\begin{equation}
4\omega^{2}\left(\frac{1}{2}-\zeta^{2}\right)\ddot{x},
\end{equation} 
is the only term in the Pais-Uhlenbeck equation \eqref{eq:PUequation} that involves the damping ratio. We might call this the ``inertial damping'' term. Interestingly, the inertial damping term vanishes for $\zeta=1/\sqrt{2}$, yielding
\begin{equation}
\ddddot{x}+\omega^{4}x=0.
\end{equation} 
This leads to a physical (rather than a purely control-theoretical) interpretation for the optimal control value. When $\zeta=1/\sqrt{2}$, the oscillator's ``jounce'' (time rate-of-change of jerk) is not affected by its current acceleration, only by its current displacement from equilibrium. In other words, there is no inertial damping for the optimal control value.

\begin{figure}[t]
 \centering
 \includegraphics[width=0.7\textwidth]{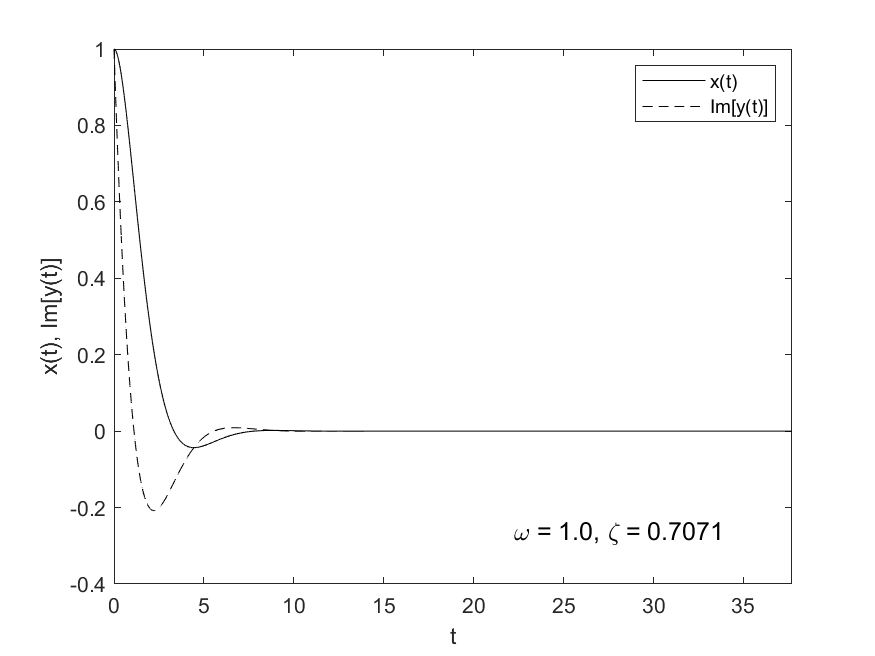}
 \caption{Representative coupled oscillator dynamics for an optimal control case with $\zeta=1/\sqrt{2}$, $\omega=1$, $x_{0}=1$, and $v_{0}=0$. The damped harmonic oscillator $x(t)$ is real-valued, while the dual oscillator $y(t)$ is imaginary.}
 \label{fig:xandyoptimalcontrol}
 \end{figure}

Additionally, $\zeta=1/\sqrt{2}$ leads to a degenerate case in which the coupled oscillator parameters $\mu_{1}=-\mu_{2}$, according to \eqref{eq:aux1}. Setting $\mu_{1}=\gamma$ and $\rho_{2}=\beta^{-2}\rho_{1}$ ($\beta\neq0$), we have
\begin{align}
\mu_{1}&=\gamma, & \mu_{2}&=-\gamma,\\
\rho_{1}&=\beta\sqrt{-\gamma^{2}-\omega^{4}}, & \rho_{2}&=\frac{1}{\beta}\sqrt{-\gamma^{2}-\omega^{4}}.
\end{align}
In Section~\ref{sec:symmetric}, when we considered the case of $\zeta=1/\sqrt{2}$, we were assuming identical masses and identical supporting springs, which amounts to setting $\gamma=0$ and $\beta=1$ here. That was the only underdamped value for which the stiffness ratio \eqref{eq:stiffnessratio} was real-valued ($r=-1$), giving
\begin{equation}
k^{*}=+im\omega^{2}, \quad k=-im\omega^{2}.
\end{equation}
Thus, optimal control is the only case in which the symmetric coupled oscillator with real masses has purely imaginary spring stiffnesses. Figure~\ref{fig:xandyoptimalcontrol} shows representative coupled oscillator dynamics for an optimal control case with $\zeta=1/\sqrt{2}$, $\omega=1$, $x_{0}=1$, and $v_{0}=0$. In this case, the damped harmonic oscillator $x(t)$ is real-valued, while the dual oscillator $y(t)$ is imaginary.

\section{Variational formulation}\label{sec:variationalformulation}

The fact that every solution to the damped harmonic oscillator equation \eqref{eq:DHOequation} also satisfies the fourth-order Pais-Uhlenbeck equation \eqref{eq:PUequation} also leads to a new variational formulation for the damped harmonic oscillator. The Pais-Uhlenbeck action is given by
\begin{equation}\label{eq:PUaction}
\mathcal{S}=\int_{t_{1}}^{t_{2}}\left[\ddot{x}^{2}-4\omega^{2}\left(\frac{1}{2}-\zeta^{2}\right)\dot{x}^{2}+\omega^{4}x^{2}\right]dt,
\end{equation}
and the Pais-Uhlenbeck equation \eqref{eq:PUequation} can be recovered by taking $\delta\mathcal{S}=0$. It follows that every solution to the damped harmonic oscillator equation \eqref{eq:DHOequation} constitutes a critical point of the Pais-Uhlenbeck action. It is instructive to compare this result to previous variational formulations of the damped harmonic oscillator. Like the dual system approach of Bateman, Morse, Feshbach, and Tikochinsky,~\cite{Bateman1931,Morse1953,Feshbach1977} the Pais-Uhlenbeck action introduces dual oscillators. The difference here is that there is actually a two-parameter family of dual oscillators $y$; moreover, the dual oscillators do not appear explicitly in the action and are not varied independently of $x$. Unlike the Kanai-Caldirola Lagrangian,~\cite{Caldirola1941,Kanai1948} the Pais-Uhlenbeck Lagrangian does not depend explicitly on time. And unlike the Caldeira-Leggett approach~\cite{Caldeira1983,Zwanzig1973,Ford1987} taken by Fukagawa and Fujitani,~\cite{Fukagawa2012} the Pais-Uhlenbeck action does not require the use of non-holonomic constraints to include the dissipative terms.

It may be that variational formulations can be found for other dissipative systems by considering higher-order equations. The problem becomes how to identify those equations. One possible method is to manipulate the second-order equation directly, as we did in Section~\ref{sec:DHOisPU}. Having arrived at \eqref{eq:PUequation}, we may multiply by $\delta x$, 
\begin{equation}
\ddddot{x}\delta x+4\omega^{2}\left(\frac{1}{2}-\zeta^{2}\right)\ddot{x}\delta x+\omega^{4}x\delta x=0,
\end{equation}
integrate by parts judiciously, 
\begin{equation}
\int_{t_{1}}^{t_{2}}\left[\ddot{x}\delta \ddot{x}-4\omega^{2}\left(\frac{1}{2}-\zeta^{2}\right)\dot{x}\delta \dot{x}+\omega^{4}x\delta x\right]dt=0,
\end{equation}
thus completing the variation and arriving at the Pais-Uhlenbeck action principle
\begin{equation}
\delta\int_{t_{1}}^{t_{2}}\left[\ddot{x}^{2}-4\omega^{2}\left(\frac{1}{2}-\zeta^{2}\right)\dot{x}^{2}+\omega^{4}x^{2}\right]dt=0.
\end{equation}

Another method is to start from the d'Alembert-Lagrange principle \eqref{eq:dAlembertLagrange} and look for a particular variation that puts it in variational form. For the damped harmonic oscillator, the d'Alembert-Lagrange principle gives
\begin{equation}
-\int_{t_{1}}^{t_{2}}(m\ddot{x}+c\dot{x}+kx)\delta{x}dt=0.
\end{equation}
Both the inertial term $m\ddot{x}\delta x$ and the spring term $kx\delta x$ can be put into variational form, because they each contain an even number of time derivatives. However, the damping term $c\dot{x}\delta x$ has an odd number of derivatives and therefore cannot be expressed as the exact variation of a work function. If we had an additional term proportional to $x\delta\dot{x}$, it could cancel out the damping term. Thus, we might look for a variational principle of the form
\begin{equation}
\int_{t_{1}}^{t_{2}}(m\ddot{x}+c\dot{x}+kx)(\delta{x}+\alpha\delta\dot{x})dt=0.
\end{equation}
However, this will give a term proportional to $\ddot{x}\delta\dot{x}$, which also has an odd number of derivatives. Thus, in general we must look for a variational principle of the form
\begin{equation}
\int_{t_{1}}^{t_{2}}(m\ddot{x}+c\dot{x}+kx)(\delta{x}+\alpha\delta\dot{x}+\beta\delta\ddot{x})dt=0.
\end{equation}
Expanding, we have
\begin{equation}
\int_{t_{1}}^{t_{2}}(m\ddot{x}\delta{x}+\alpha m\ddot{x}\delta\dot{x}+\beta m\ddot{x}\delta\ddot{x}+c\dot{x}\delta{x}+\alpha c\dot{x}\delta\dot{x}+\beta c\dot{x}\delta\ddot{x}+kx\delta{x}+\alpha kx\delta\dot{x}+\beta kx\delta\ddot{x})dt=0.
\end{equation}
The terms with odd numbers of derivatives are $\alpha m\ddot{x}\delta\dot{x}$, $c\dot{x}\delta{x}$, $\beta c\dot{x}\delta\ddot{x}$, and $\alpha kx\delta\dot{x}$. Upon integration by parts, we want $\alpha m\ddot{x}\delta\dot{x}$ to cancel $\beta c\dot{x}\delta\ddot{x}$, and for $c\dot{x}\delta{x}$ to cancel $\alpha kx\delta\dot{x}$. This requires
\begin{equation}
\alpha m=\beta c \quad \text{and} \quad c=\alpha k,
\end{equation}
which gives $\alpha=c/k$ and $\beta=m/k$. Hence, we are looking for a variational principle of the form
\begin{equation}
\int_{t_{1}}^{t_{2}}(m\ddot{x}+c\dot{x}+kx)\left(\delta{x}+\frac{c}{k}\delta\dot{x}+\frac{m}{k}\delta\ddot{x}\right)dt=0.
\end{equation}
We observe that $\delta{x}+(c/k)\delta\dot{x}+(m/k)\delta\ddot{x}$ is actually a variation of the governing equation, up to a factor of $k$. Expanding, we have
\begin{equation}
\int_{t_{1}}^{t_{2}}\left(m\ddot{x}\delta{x}+\frac{mc}{k} \ddot{x}\delta\dot{x}+\frac{m^{2}}{k}\ddot{x}\delta\ddot{x}+c\dot{x}\delta{x}+\frac{c^{2}}{k} \dot{x}\delta\dot{x}+\frac{mc}{k}\dot{x}\delta\ddot{x}+kx\delta{x}+cx\delta\dot{x}+mx\delta\ddot{x}\right)dt=0.
\end{equation}
Judicious integration by parts yields
\begin{equation}
\int_{t_{1}}^{t_{2}}\left(-m\dot{x}\delta\dot{x}+\frac{mc}{k} \ddot{x}\delta\dot{x}+\frac{m^{2}}{k}\ddot{x}\delta\ddot{x}+c\dot{x}\delta{x}+\frac{c^{2}}{k} \dot{x}\delta\dot{x}-\frac{mc}{k}\ddot{x}\delta\dot{x}+kx\delta{x}-c\dot{x}\delta{x}-m\dot{x}\delta\dot{x}\right)dt=0,
\end{equation}
which simplifies to
\begin{equation}
\int_{t_{1}}^{t_{2}}\left[\frac{m^{2}}{k}\ddot{x}\delta\ddot{x}+\left(\frac{c^{2}}{k} -2m\right)\dot{x}\delta\dot{x}+kx\delta{x}\right]dt=0.
\end{equation}
Each term can now be put into variational form, yielding
\begin{equation}
\delta\int_{t_{1}}^{t_{2}}\left[\frac{m^{2}}{2k}\ddot{x}^{2}+\frac{1}{2}\left(\frac{c^{2}}{k} -2m\right)\dot{x}^{2}+\frac{1}{2}kx^{2}\right]dt=0,
\end{equation}
or equivalently,
\begin{equation}
\delta\int_{t_{1}}^{t_{2}}\left[\ddot{x}^{2}-4\omega^{2}\left(\frac{1}{2}-\zeta^{2}\right)\dot{x}^{2}+\omega^{4}x^{2}\right]dt=0.
\end{equation}
Future work may consider whether either of the above approaches generalizes to other kinds of dissipative systems.

\section{Summary and conclusion}\label{sec:conclusion}

The main result of this paper is that the classical damped harmonic oscillator belongs to the family of fourth-order Pais-Uhlenbeck oscillators. Three immediate corollaries follow:
\begin{enumerate}[(I)]
  \item For every damped harmonic oscillator $x$, there exists a two-parameter family of dual oscillators $y$ satisfying the Pais-Uhlenbeck equation. The damped harmonic oscillator and any of its duals constitute a system of two coupled oscillators with non-standard spring stiffnesses. The various cases are summarized in Table~\ref{tab:cases} for coupled oscillators with identical masses and identical supporting springs. Of particular interest is the case of overdamping $\zeta>1$, for which the stiffness of the coupling spring is positive while the stiffnesses of the supporting springs are negative. This scenario should be physically achievable~\cite{Lee2007} and may even find applications to vibration suppression systems.
  \item The optimal control value $\zeta=1/\sqrt{2}$, which gives the best tradeoff between rise time and percent overshoot in response to a step input,~\cite{Jacobs1965} appears prominently in the fourth-order formulation and has a clear physical interpretation. For this value of the damping ratio, there is no inertial damping.
  \item The solutions of the damped harmonic oscillator make the Pais-Uhlenbeck action stationary, yielding a new variational formulation for the damped harmonic oscillator. Two systematic methods have been given for arriving at the Pais-Uhlenbeck action starting from the damped harmonic oscillator equation. It may be possible to generalize these approaches and identify stationary actions for other kinds of dissipative systems which do not conform to Hamilton's principle.
\end{enumerate}

\section*{Declarations}

\subsection*{Conflicts of interest}

The author declares that he has no conflicts of interest or competing interests.
%
%
%
\subsection*{Data availability}

Data sharing is not applicable to this article as no new data were created or analyzed in this study.
%
%

\bibliography{Sanders2021_DHO}

\end{document}